\pdfoutput=1	

\documentclass[10pt]{iopart}

\usepackage{hyperref}
\usepackage{graphicx}

\begin{document}

\title[Metamaterials for light rays]{Metamaterials for light rays:  ray optics without wave-optical analog in the ray-optics limit}

\author{Alasdair C.\ Hamilton and Johannes Courtial}
\ead{j.courtial@physics.gla.ac.uk}
\address{Department of Physics and Astronomy, Faculty of Physical Sciences, University of Glasgow, Glasgow G12~8QQ, United~Kingdom}

\date{\today}

\begin{abstract}
Volumes of sub-wavelength electromagnetic elements can act like homogeneous materials: metamaterials. In analogy, sheets of optical elements such as prisms can act ray-optically like homogeneous sheet materials.
In this sense, such sheets can be considered to be metamaterials for light rays (METATOYs).
METATOYs realize new and unusual transformations of the directions of transmitted light rays.
We study here, in the ray-optics and scalar-wave limits, the wave-optical analog of such transformations, and we show that such an analog does not always exist.
Perhaps this is the reason why  many of the ray-optical possibilities offered by METATOYs have never before been considered.
\end{abstract}

\submitto{\NJP}


\section{Introduction}
Metamaterials \cite{Smith-et-al-2004} allow unprecedented control over light waves.
This has led to the development of new concepts, such as the coordinate-transform design paradigm \cite{Pendry-et-al-2006}.
Metamaterials consist of sub-wavelength-size wave-optical elements, namely resonant (for a particular frequency) electro-magnetic circuits, filling a volume.
This structure gives them the wave-optical properties of a homogeneous material with macroscopic parameters, namely the permittivity, $\epsilon$, and the permeability, $\mu$~\cite{Smith-et-al-2004}.

\begin{figure}
\centering \includegraphics{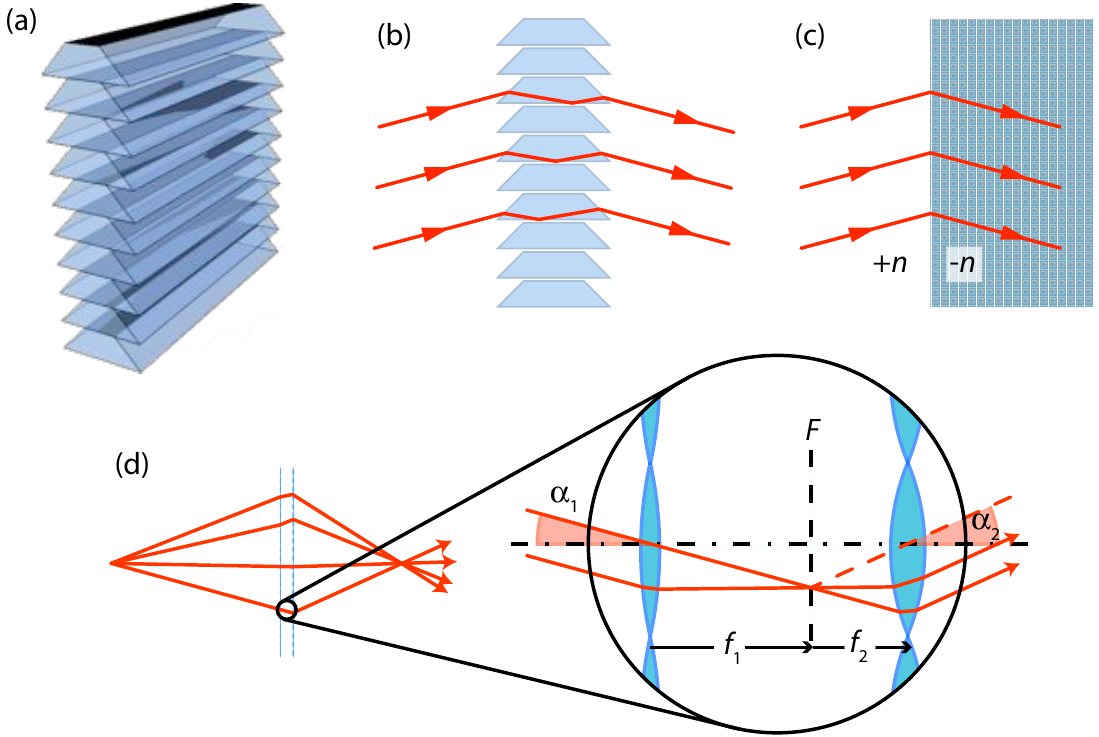}
\caption{\label{METATOYs-figure}
Examples of METATOYs.
(a)~An array of stacked, miniaturized, Dove prisms forms a Dove-prism sheet \cite{Courtial-Nelson-2008}.
For illustration purposes, the Dove-prism sheet shown here contains only very few (10) Dove prisms.
In our simulations (see Fig.\ \ref{views-figure}(b) and (c)), the number of Dove prisms per Dove-prism sheet typically varies between 100 and 1000.
(b)~Transmission through a Dove-prism sheet flips one ray-direction component, in the case shown here the vertical component.
As far as this flipped ray-direction component is concerned, the effect is negative refraction at the interface between media with opposite refractive indices~(c).
(d)~Two parallel arrays of miniaturized lenses (``lenslet arrays'') that share the same focal plane, $F$, (that is, they are confocal) refract light rays approximately like the interface between transparent media whose refractive indices, $n_1$ and $n_2$, have a ratio $n_2 / n_1 \approx \eta = -f_2 / f_1$, where $f_1$ and $f_2$ are the respective focal lengths of the two lenslet arrays \cite{Courtial-2008a}.}
\end{figure}

Motivated by the desire to build optical elements that have some of the \emph{visual} properties of metamaterials on an everyday size scale and across the entire visible wavelength spectrum, we recently started to investigate sheets formed by miniaturized optical elements that change the direction of transmitted light rays (Fig.\ \ref{METATOYs-figure}).
In the simplest case (which we consider here), these sheets are constructed to work in air, such that each light ray enters and exits the sheet on opposite sides with the same intensity and without being offset, and that the change in light-ray direction is independent of the position where the ray hits the sheet.
In reality, the sheets introduce a number of imperfections.
They do, for example, offset transmitted light rays, but the offsets are limited to the size of the individual elements' aperture diameter, which can be miniaturized, in principle until ray optics breaks down (10s of wavelengths).
Nevertheless, a sheet's visual properties, and generally its ray-optical behavior, is well described by macroscopic parameters that characterize the ray-direction change.
Like metamaterials, such sheets ``are materials with designed properties that stem from structure, not substance'' \cite{Leonhardt-Philbin-2006}, 
so we call them \emph{\underline{meta}ma\underline{t}erials f\underline{o}r ra\underline{y}s (METATOYs)}\footnote{Note that different structured surfaces have previously been described as metamaterials \cite{Leskova-et-al-2007}.}.
Because the ray-direction change in METATOYs is a generalization of refraction, we call it \emph{meta-refraction}.

\begin{figure}[t]
\centering \includegraphics{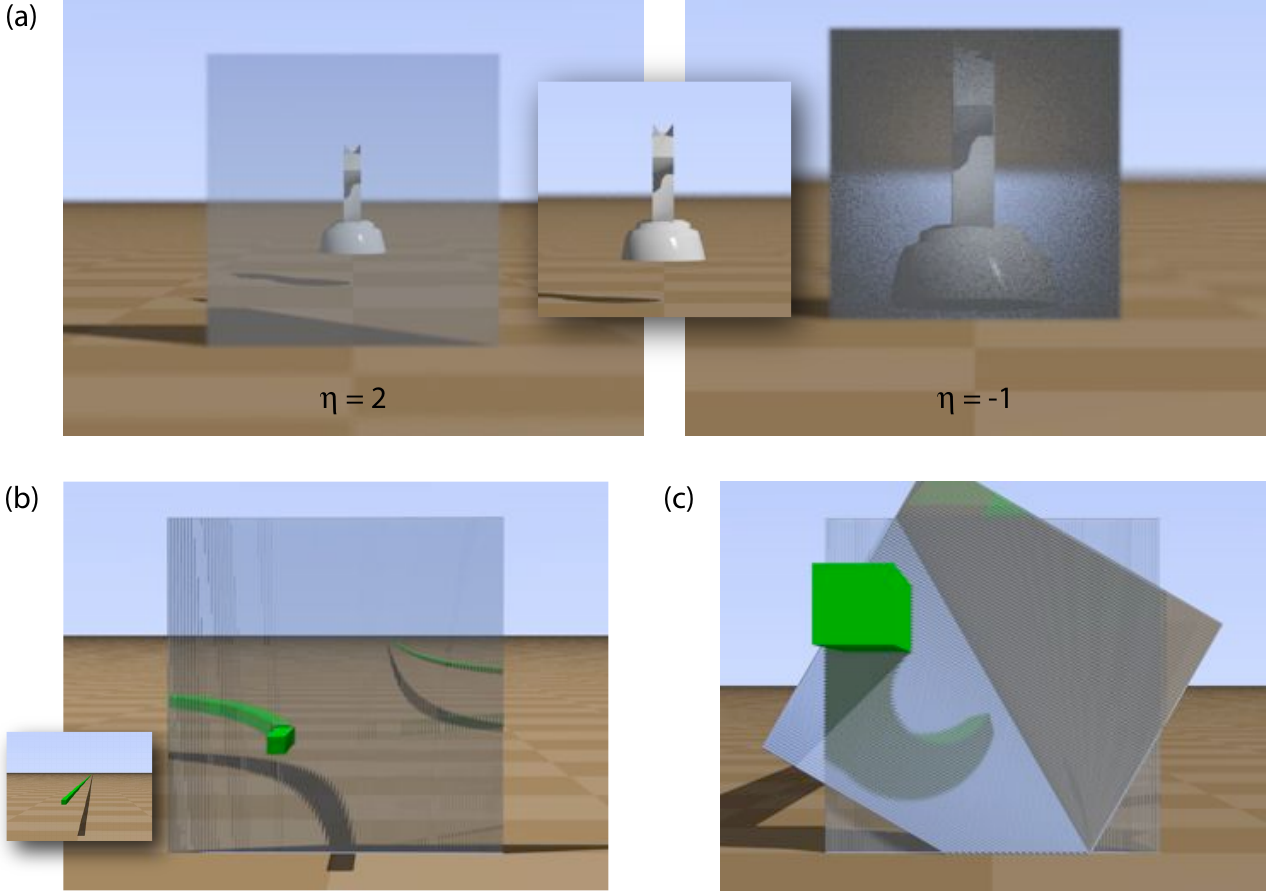}
\caption{\label{views-figure}
Example of views through METATOYs.
(a)~When an object is seen through confocal lenslet arrays (Fig.\ \ref{METATOYs-figure}(d)), the distance between the arrays and the object appears to be stretched by a factor $\eta$.
The two examples show a stationary chess piece (see inset) seen through confocal lenslet arrays with $\eta = 2$ and $\eta = -1$, respectively making the chess piece appear further away and in front of the arrays.
(The latter case is the imaging case discussed by Pendry \cite{Pendry-2000} for negative refraction between media with opposite refractive indices, but without restoration of the evanescent-wave components).
(b)~When seen through a Dove-prism sheet (Fig.\ \ref{METATOYs-figure}(a)), a straight line perpendicular to the sheet appears bent into a hyperbola \cite{Hamilton-Courtial-2008a}.
Here the straight line is approximated by a thin green cuboid, shown on its own for comparison in the inset.
(c)~When seen through a ray-rotation sheet \cite{Hamilton-et-al-2009}, a cuboid appears twisted.
Part of the cuboids in (b) and (c) extend in front of the sheet so that each cuboid's geometry can be appreciated.
All views are ray-tracing simulations through the detailed sheet structure, performed with the free software POV-Ray \cite{POV-Ray}.
Details of the simulations shown in (a) and (b) can be found in the references given.
The ray-rotation sheet in (c) comprises two Dove-prism sheets \cite{Hamilton-Courtial-2008a}, each containing 100 Dove prisms, that are rotated with respect to each other through $60^\circ$, together forming a sheet that rotates the direction of transmitted light rays through $120^\circ$ around the local sheet normal.}
\end{figure}

Meta-refraction can be familiar, like in confocal lenslet arrays that act approximately like the interface between optical media with different positive or negative refractive indices (Fig.\ \ref{views-figure}(a))
\cite{Courtial-2008a,Courtial-Nelson-2008}.
It can also be very unfamiliar, like in Dove-prism structures (Fig.\ \ref{views-figure}(b)) \cite{Hamilton-Courtial-2008a}, especially those that rotate the light-ray direction around the local sheet normal (Fig.\ \ref{views-figure}(c))
\cite{Hamilton-et-al-2009}.
Combinations of such ray-rotation sheets can have unusual imaging properties \cite{Hamilton-Courtial-2008c}, and a Fermat's-principle treatment of light-ray rotation around the local sheet normal leads to a formulation in terms of an interface between complex refractive indices~\cite{Sundar-et-al-2009}.

Here we begin to investigate the wave optics of METATOYs.
We investigate whether every light-ray field that has been transformed by a METATOY has a simplified wave-optical analog:
a complex scalar field whose intensity is that of the light-ray field, and whose phase gradient points in the light-ray direction.
Our definition uses two approximations:  the description of light by a single scalar field, which is customary in a number of areas of optics (e.g.\ \cite{Berry-1998a}), and the (arguably appropriate) ray-optics limit~\cite{Landau-Lifshitz-II-light-rays}.

Note that we are not calculating the light-wave field behind a METATOY, which is, of course, always possible.
Instead, we ask the question if a simplified -- but, within the scope of the simplifications, perfect (in the sense explained in section \ref{wave-optical-analog-section}) -- wave-optical analog of METATOYs can exist, \emph{even in principle}.

This paper is organized as follows.
In section \ref{wave-optical-analog-section} we define the wave-optical analog of a light-ray field, and we find a requirement for this analog to exist.
By studying a simple example in section \ref{example-section}, we then construct light-ray fields without wave-optical analog.
This allows us to formulate the condition for the existence of a wave-optical analog in terms of topological-charge density (section \ref{topological-charge-section}).
Section \ref{mapping-section} outlines the relationship between our work and geometric transformations that cannot be implemented exactly with continuous holograms.
In section \ref{analog-of-metarefraction-section} we then study briefly the conditions for particular types of meta-refraction to turn any light-ray field with wave-optical analog into another such field, which is the condition for the type of meta-refraction to have a wave-optical analog.
Finally, we provide conclusions.

\section{\label{wave-optical-analog-section}Wave-optical analog of light-ray fields}
A light-ray field is fully described by a three-dimensional field of normalized three-dimensional direction vectors and an intensity field.
For our purposes it is sufficient to consider the transverse ($x$ and $y$) direction components in the $z=0$ plane, which we describe by the vector field
\begin{equation}
\bi{r}(x,y) = \left( \begin{array}{c} r_x(x,y) \\ r_y(x,y) \end{array} \right).
\label{ray-field-equation}
\end{equation}
We restrict ourselves here to fields in which, wherever the intensity is non-zero, the light-ray directions vary smoothly, by which we mean that the partial derivatives $\partial_x r_x$, $\partial_y r_x$, $\partial_x r_y$, $\partial_y r_y$ exist (we use the notation $\partial_a b \equiv \partial b / \partial a$).

We define the wave-optical analog of such a light-ray field as follows.
The analogous monochromatic light wave, if it exists, is a complex field with a cross-section in the transverse plane of the form
\begin{equation}
u(x,y) = \sqrt{I(x,y)} \exp\left(\rmi \phi(x,y)\right).
\label{complex-field-equation}
\end{equation}
We want this field to be ``non-pathological''; we ask for it to be continuous and differentiable everywhere -- the usual properties asked from a physical field.
Wherever the intensity $I$ is non-zero, the phase $\phi$ then has to be continuous and differentiable.
We translate the ray field's brightness and direction into the complex field's intensity and three-dimensional phase gradient, $\nabla \phi$, respectively.
The latter choice follows the usual ray-optics limit \cite{Landau-Lifshitz-II-light-rays}.

We have not yet specified the magnitude of the phase gradient.
In a monochromatic plane wave of wavelength $\lambda$, the magnitude of the phase gradient is $2 \pi/\lambda$, so the relationship between the phase gradient and the (normalized) ray direction is
\begin{equation}
\nabla \phi(x,y) = \frac{2 \pi}{\lambda} \bi{r}(x,y).
\label{local-plane-wave-scenario-equation}
\end{equation}
(Note that $\phi(x,y)$ and $\bi{r}(x,y)$ are defined such that this equation describes the $x$ and $y$ components of the three-dimensional phase gradient and normalized light-ray direction, multiplied by $2 \pi/\lambda$, in the $z=0$ plane.)
This is also generally the case in the ray-optics limit, in which the wavelength is so short that, on the scale of a few wavelengths, any wave locally looks planar.
This is the main scenario we consider here.

Sometimes we also venture slightly further from the ray-optics limit by considering a scenario in which the magnitude of the phase gradient is not $2 \pi/\lambda$. 
This occurs routinely in general monochromatic waves:
in optical superoscillations, for example, the magnitude of the phase gradient can be greater, even arbitrarily large, but conventional wisdom has it that this can only happen when the intensity gets very close to zero \footnote{This particular ``conventional wisdom'' is a generalisation of special cases such as Ref.\ \cite{Berry-1994}, and is correct in a statistical sense \cite{Dennis-et-al-2008}.},
and in special cases such as standing waves the phase gradient can be zero.
In this ``less ray-optical scenario'' we therefore allow the phase gradient to be of any length, as long as it has a direction (i.e.\ length $\neq 0$), is finite, and does not point in the opposite direction from the light-ray direction \footnote{In negative-refractive-index materials, the wave vector points in the opposite direction of the ``propagation direction'' \cite{Smith-et-al-2004}.
We consider here only METATOYs embedded in a positive-refractive-index material.
We also restrict ourselves to homogeneous materials.}:
\begin{equation}
\nabla \phi(x,y) = f(x,y) \frac{2 \pi}{\lambda} \bi{r}(x,y),
\label{less-ray-optical-scenario-equation}
\end{equation}
where $0 < f(x,y) < \infty$.

We now consider a METATOY immediately in front of the $z=0$ plane, and concentrate on the field immediately behind the METATOY.
The existence of the partial derivatives of the components of $\bi{r}(x,y)$, specifically $\partial_y r_x$ and $\partial_x r_y$, implies that the mixed partial derivatives of $\phi(x,y)$, $\partial_x \partial_y \phi$ and $\partial_y \partial_x \phi$, exist.
Under these (and, indeed, less strict)
conditions the theorem on the order of differentiation \cite{Courant-1936-II-change-of-differentiation-order} holds, stating that
\begin{equation}
\partial_y \partial_x \phi = \partial_x \partial_y \phi.
\label{Schwarz-theorem}
\end{equation}
Using Eqn (\ref{local-plane-wave-scenario-equation}), this can be translated into the following condition on the transverse ray directions, $r_x$ and $r_y$:
\begin{equation}
\partial_x r_y = \partial_y r_x.
\label{Schwarz-for-rays-equation}
\end{equation}
In the less ray-optical scenario, the corresponding condition is
\begin{equation}
\partial_y \left[ f(x,y) r_x(x,y) \right] = \partial_x \left[ f(x,y) r_y(x,y) \right].
\label{Schwarz-for-rays-equation-2}
\end{equation}
If a light-ray field does not satisfy Eqn (\ref{Schwarz-for-rays-equation}) (or, in the case of the less ray-optical scenario, Eqn (\ref{Schwarz-for-rays-equation-2}) for at least one function $f(x,y) > 0$) wherever the intensity is non-zero,
then the phase $\phi(x,y)$, and with it the complex field $u(x,y)$ that corresponds to the ray directions $r_x(x,y)$ and $r_y(x,y)$, cannot exist there.
In this sense, Eqns (\ref{Schwarz-for-rays-equation}) and (\ref{Schwarz-for-rays-equation-2}) are therefore in the different scenarios necessary conditions for the existence of a light-ray field's wave-optical analog.

\section{\label{example-section}Example}
As an example, we consider here a specific light beam passing through a METATOY immediately in front of the $z=0$ plane that rotates the light-ray direction through $90^\circ$ around the $z$ direction without offsetting the ray position.
Such local ray rotation can be achieved with
combinations of Dove-prism sheets~\cite{Hamilton-et-al-2009}.

We consider a light-ray field which has, in the plane immediately in front of the ray-rotating METATOY, the following properties.
In the neighbourhood of the point $(x,y) = (0,0)$, the brightness is uniform and the light-ray directions are given by the gradient of the quadratic phase function
\begin{equation}
\phi_0(x,y) = x^2,
\label{example-phase-equation}
\end{equation}
so the  $x$ and $y$ components of the ray-direction field are
\begin{equation}
\bi{r}_0(x,y) = \nabla \phi_0(x,y) = \left(\begin{array}{c} 2 x \\ 0 \end{array}\right).
\end{equation}
After passage through the METATOY, all ray directions have been rotated through $90^\circ$ around the $z$ axis, so in the plane immediately behind the METATOY the ray-direction components are
\begin{equation}
\bi{r}(x,y) = \left(\begin{array}{c} 0 \\  2 x \end{array}\right).
\label{rotated-directions-equation}
\end{equation}

Can such a light-ray-direction field have a wave-optical analog?
The $x$ and $y$ components of $\bi{r}$ clearly violate the condition for the existence of a wave-optical analog in the ray-optics limit: substitution into Eqn (\ref{Schwarz-for-rays-equation}) gives ``$2 = 0$''.
In the less-ray-optical scenario, the situation is slightly more complicated:
substituting the $x$ and $y$ components of $\bi{r}$ into Eqn (\ref{Schwarz-for-rays-equation-2}), we get 
\begin{equation}
0 = \partial_x \left( f(x,y) 2 x \right).
\end{equation}
This condition can only be satisfied if
\begin{equation}
f(x,y) = \frac{g(y)}{x},
\end{equation}
where $g(y)$ is function of $y$ that is greater than zero (as $f(x,y) > 0$).
It is clear that, at $x=0$, $f(x,y)$ diverges to infinity and changes sign at $x=0$, contrary to the conditions of the less ray-optical scenario.
In either scenario the light-ray field given by Eqn (\ref{rotated-directions-equation}) therefore has no wave-optical analog.

\begin{figure}
\begin{center}
\includegraphics{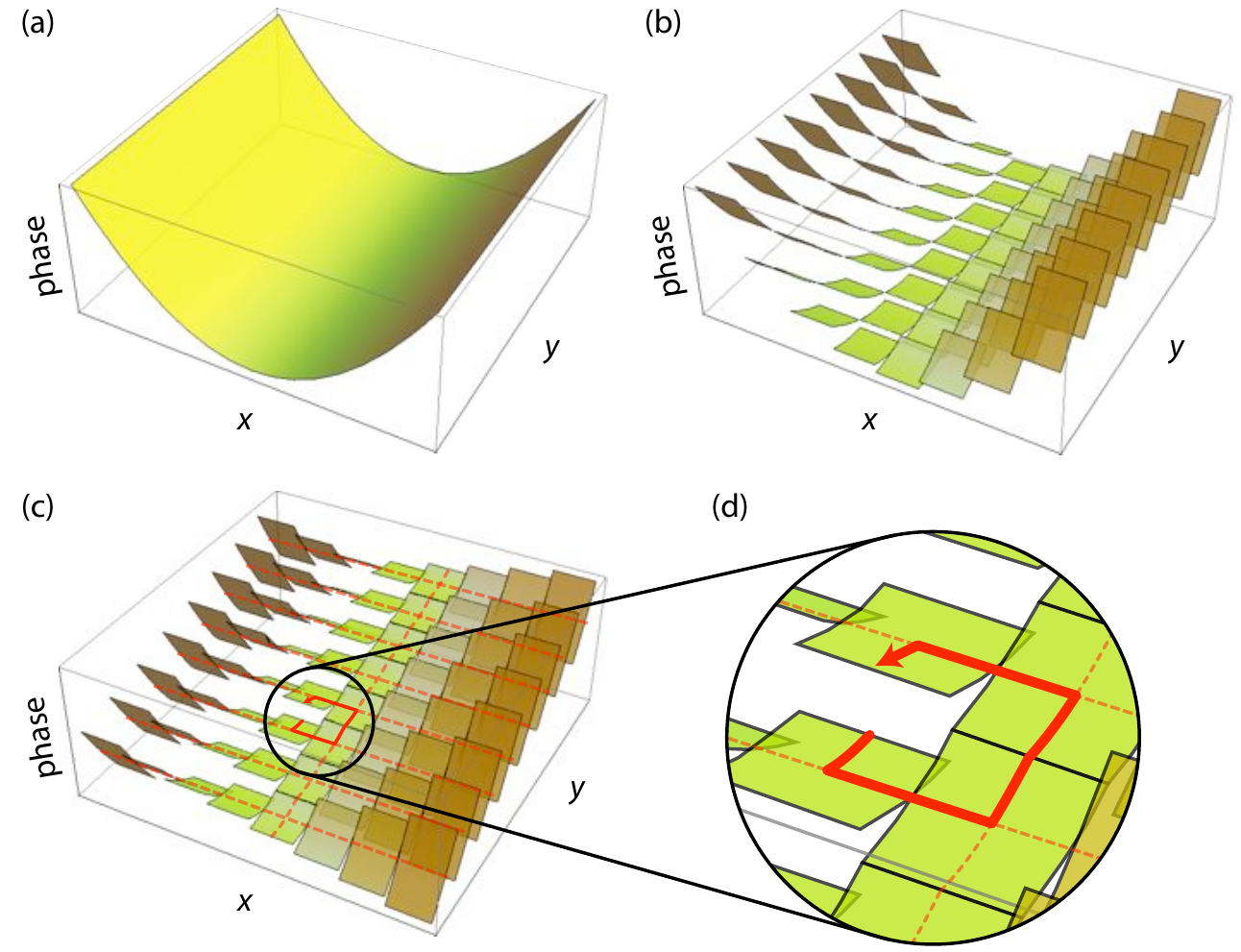}
\end{center}
\caption{\label{piecewise-rotation-figure}Piecewise rotation of the phase.
(a)~Plot of the phase $\phi_0(x,y) = x^2$.
(b)~$\phi_0(x,y)$, piecewise rotated through $90^\circ$.
(c)~The rotated pieces are individually phase-shifted such that they fit together along the dashed lines (red in the online version).
(d)~The thick solid arrow (red online) indicates a non-zero path integral along a closed loop in the $z=0$ plane.}
\end{figure}

To illustrate the lack of a wave-optical analog in the ray-optics limit in this example, we consider a piecewise rotation of the phase.
Figure \ref{piecewise-rotation-figure}(a) shows the quadratic phase function $\phi_0(x,y)$ described by Eqn (\ref{example-phase-equation}).
Figure \ref{piecewise-rotation-figure}(b) shows the result of dividing the $z=0$ plane, and with it $\phi_0(x,y)$, into squares of equal size and rotating each square through $90^\circ$ around its center.
Such piecewise phase rotation is an approximation to truly local light-ray rotation and an idealization \footnote{Dove-prism ray-rotation sheets do not rotate the ray direction in square pieces, but they mirror the ray direction in stripes, twice, whereby the stripes have a different orientation; they do not even mirror the phase in the stripes perfectly; and they change not only the phase, but the entire complex field.} of the effect of passage through local-light-ray rotators based on Dove-prism sheets \cite{Hamilton-et-al-2009}.

The question then is whether or not it is possible to construct a continuous phase function from the rotated pieces of $\phi(x,y)$ by moving the rotated pieces individually up or down.
(We are not concerned here with the mechanism by which such a spatially-varying phase shift might happen; all we want to know is whether or not a phase function with the derivatives corresponding to the locally rotated ray-direction field can exist.)
Figure \ref{piecewise-rotation-figure}(c) shows the phase after all the pieces have been vertically shifted such that they fit together along the dashed lines. 
It is immediately clear that the resulting phase function is still full of discontinuities along at least two of each piece's edges, irrespective of the size of the pieces.

If we consider the case of rotation of the phase gradient at every point as a similar kind of piecewise rotation in the limit of infinitely small pieces, we find that discontinuities along at least one edge of each piece of the phase translate into discontinuities which are infinitely small (they scale linearly with the size of the pieces), but which exist at \emph{every point} of the resulting phase function.
Such a phase function can be continuous everywhere, but it is not differentiable anywhere, and therefore it does not have an associated ray direction according to Eqn~(\ref{local-plane-wave-scenario-equation}).

Mathematically, 
an effect of the pieces not fitting together like in Fig.\ \ref{piecewise-rotation-figure}(c) is that the path integral of the phase gradient along closed paths, for example the path shown in Fig.\ \ref{piecewise-rotation-figure}(d), is non-zero.
The closed-loop path integral around an infinitesimally small square of side length $\delta$, centred at $(x_0,y_0)$, can be written in the form
\begin{eqnarray}
\Delta \phi &=& \delta \left(
	\left. \partial_x \phi \right|_{(x_0,y_0-\delta/2)} +
	\left. \partial_y \phi \right|_{(x_0+\delta/2,y_0)} -
\right. \nonumber \\
&& \left.
	\left. \partial_x \phi \right|_{(x_0,y_0+\delta/2)} -
	\left. \partial_y \phi \right|_{(x_0-\delta/2,y_0)}
\right);
\label{path-integral}
\end{eqnarray}
the condition for the path integral to be zero is
\begin{equation}
\Delta \phi = 0.
\label{path-integral-zero-condition}
\end{equation}
The partial derivatives in equation (\ref{path-integral}) can be linearized to become
\begin{equation}
\left. \partial_x \phi \right|_{\left( x_0,y_0+\delta/2 \right)} =
\left. \partial_x \phi \right|_{\left( x_0,y_0-\delta/2 \right)} + 
\delta \left. \partial_x \partial_y \phi \right|_{(x_0,y_0)}
\end{equation}
and
\begin{equation}
\left. \partial_y \phi \right|_{\left( x_0+\delta/2,y_0 \right)} =
\left. \partial_y \phi \right|_{\left( x_0-\delta/2,y_0 \right)} + 
\delta \left. \partial_y \partial_x \phi \right|_{(x_0,y_0)};
\end{equation}
Eqn (\ref{path-integral-zero-condition}) then becomes
\begin{equation}
\left. \partial_y \partial_x \phi \right|_{(x_0,y_0)} = 
\left. \partial_x \partial_y \phi \right|_{(x_0,y_0)}.
\label{Schwarz-at-xy0}
\end{equation}
We have therefore derived Eqn (\ref{Schwarz-theorem}) as a necessary condition for the existence of a function $\phi(x,y)$ which is continuous and differentiable at position $(x_0, y_0)$.
As we want $\phi(x,y)$ to have these properties wherever the intensity is non-zero, Eqn (\ref{Schwarz-theorem}) has to hold wherever this is the case.

\section{\label{topological-charge-section}Formulation in terms of topological-charge density}
The closed-loop path integral (\ref{path-integral}) can be interpreted as enclosing a topological charge (phase change divided by $2 \pi$) of
\begin{equation}
\Delta m = \frac{\Delta \phi}{2 \pi} = \frac{\delta^2}{2 \pi} \left( \partial_x r_y - \partial_y r_x \right).
\end{equation}
As the area enclosed by the path is $\delta^2$,  we can define the topological charge density (topological charge per area):
\begin{equation}
\mu = \frac{\Delta m}{\delta^2} = \frac{\partial_x r_y - \partial_y r_x}{2 \pi}.
\label{topological-charge-density}
\end{equation}
The condition on light-ray fields to have a wave-optical analog can then be formulated in terms of this topological-charge density:
Eqn (\ref{Schwarz-for-rays-equation}), which has to hold wherever the intensity is non-zero, becomes $\mu = 0$.
At those points where the topological-charge density is not zero, the light intensity has to be zero.
This means that the product of topological-charge density and intensity, a quantity closely related to the density of optical orbital angular momentum \cite{Allen-et-al-1992}, has to be zero everywhere:
\begin{equation}
\mu I = 0.
\label{mu-I-equation}
\end{equation}

\section{\label{mapping-section}Relationship with holographic geometric mappings}

Our work is closely related to work on phase holograms that map the intensity cross-section in the hologram plane of a collimated light beam into the hologram's Fourier plane \cite{Sandstrom-Lee-1983,Paterson-1994}.
Interesting mappings include rotation of the entire intensity pattern around a point, which requires an everywhere-discontinuous phase pattern that contradicts Eqn (\ref{Schwarz-theorem}).

In practice, such transformations have been approximated with holograms that concentrate the discontinuities.
One approach \cite{Paterson-1994} concentrates the discontinuities in points with integer topological charge (optical vortices).
A similar approach could perhaps be applied to metarefraction, leading to an alternative type of METATOYs.
In an alternative approach \cite{Sandstrom-Lee-1983}, the phase is piecewise continuous, and the discontinuities are concentrated in lines (the edges of the pieces).
This latter approach is analogous to the piecewise phase rotation discussed in section~\ref{example-section}.

\section{\label{analog-of-metarefraction-section}Wave-optical analog of meta-refraction}
The conditions for which a type of meta-refraction has a wave-optical analog are those under which Eqn (\ref{Schwarz-for-rays-equation}) is satisfied by any light-ray field resulting from meta-refraction of an incident light-ray field that satisfies Eqn~(\ref{Schwarz-for-rays-equation}).

As our example we use ray rotation again.
The incident light-ray field
\begin{equation}
\bi{r}(x,y) = \left( \begin{array}{c} r_x \\ r_y \end{array} \right)
\end{equation}
corresponds to a wave-optical field, so it satisfies Eqn (\ref{Schwarz-for-rays-equation}).
The rotated light-ray field is
\begin{equation}
\left(\begin{array}{c} r^{(\alpha)}_x \\ r^{(\alpha)}_y \end{array}\right) =
\left(\begin{array}{c} r_x \cos \alpha - r_y \sin \alpha \\ r_x \sin \alpha + r_y \cos \alpha \end{array}\right).
\end{equation}
It has a wave-optical analog if it satisfies Eqn (\ref{Schwarz-for-rays-equation}) with $r_x$ and $r_y$ replaced by $r^{(\alpha)}_x$ and $r^{(\alpha)}_y$, respectively:
\begin{equation}
\partial_x r_x \sin \alpha + \partial_x r_y \cos \alpha =
\partial_y r_x \cos \alpha - \partial_y r_y \sin \alpha.
\label{Schwarz-for-rotated-rays}
\end{equation}
Because the incident light-ray field satisfies Eqn (\ref{Schwarz-for-rays-equation}), all terms proportional to $\cos \alpha$ cancel, so the condition becomes
\begin{equation}
\left( \partial_x r_x + \partial_y r_y \right) \sin \alpha = 0.
\label{Schwarz-for-rotated-rays-2-equation}
\end{equation}
This means local ray rotation through an angle $\alpha$ can have a wave-optical analog that works for any input field, and therefore any arbitrary functions $r_x(x,y)$ and $r_y(x,y)$, only if $\sin \alpha = 0$, which is the case for rotation angles $0^\circ$ and $180^\circ$.
The rotation angle $0^\circ$ trivially corresponds to no refraction at all, and
the rotation angle $180^\circ$ formally corresponds to negative refraction in which the refractive index changes sign, but not magnitude ($n_1 = -n_2$) \footnote{Note, however, that genuine negative refraction produces a field in which the phase gradient points in the direction opposite \cite{Pendry-Smith-2004} to that due to $180^\circ$ rotation around the $z$ direction of the local phase gradient.}.
For other angles, the rotated light-ray field can still have a wave-optical analog, provided it satisfies Eqn~(\ref{Schwarz-for-rotated-rays-2-equation}).

\section{Conclusions}
We argue here that METATOYs are ray-optical analogs of metamaterials, and that they can produce visual effects best described by ray-optical transformations that have no simple wave-optical analog.
This might be the reason why such transformation have never before been considered, let alone realized.

We have restricted ourselves here to the ray-optics limit of wave optics, as this is easiest to tackle.
A natural continuation of this work would be a full, non-scalar, wave-optics treatment that translates light-ray direction into a suitable quantity like the Poynting vector.

\ack
Thanks to Stephen Barnett, Mark Dennis, Kevin O'Holleran and Carl Paterson for very helpful discussions.
ACH is supported by the UK's Engineering and Physical Sciences Research Council (EPSRC).
JC is a Royal Society University Research Fellow.

\section*{References}

\bibliographystyle{osajnl}
\bibliography{/Users/johannes/Documents/work/library/Johannes}

\end{document}